%% file: main.tex
\documentclass[10pt,conference]{IEEEtran} 
\IEEEoverridecommandlockouts

\usepackage{cite}
\usepackage{amsmath,amssymb,amsfonts}
\usepackage{algorithmic}
\usepackage{graphicx}
\usepackage{textcomp}
\usepackage{xcolor}
\usepackage{fontawesome}
\usepackage{multirow}
\usepackage{colortbl}
\usepackage{soul}
\usepackage{enumitem}
\usepackage{fp}
\usepackage[most]{tcolorbox}
\usepackage{tikz}
\usetikzlibrary{positioning, arrows.meta, calc}
\usepackage{booktabs}
\usepackage{tabularx}
\usepackage{adjustbox}
\usepackage{array}
\usepackage{makecell}
\usepackage{url}
\usepackage{xurl}      
\usepackage[hidelinks]{hyperref}  
\def\BibTeX{{\rm B\kern-.05em{\sc i\kern-.025em b}\kern-.08em
    T\kern-.1667em\lower.7ex\hbox{E}\kern-.125emX}}

\newcommand{\boldification}[1]{\ifdraft\indent ** \textbf{#1}** \\ \indent\else\relax\fi}
\newif\ifdraft
 \draftfalse 

\definecolor{bubbleLight}{HTML}{FEF9EA} 
\definecolor{bubbleMid}{HTML}{E6B582}   
\definecolor{bubbleDark}{HTML}{D99A5F}  
\definecolor{bubbleText}{HTML}{2B2B2B}

\definecolor{Reddit}{HTML}{FFE6C8}
\definecolor{Blog}{HTML}{FFD9AD}
\definecolor{MML}{HTML}{FFC985}

\newtcolorbox{bubble}[2][]{%
  enhanced,
  breakable,
  colback=bubbleLight!28!white,
  colframe=bubbleDark,
  boxrule=0.8pt,
  arc=1.5mm,
  width=\linewidth,
  boxsep=1pt,
  top=3pt,
  bottom=3pt,
  left=3pt,
  right=3pt,
  before skip=5pt,
  after skip=5pt,
  fonttitle=\bfseries\color{black},
  coltitle=white,
  colbacktitle=bubbleMid,
  title=#2,
  #1
}

\definecolor{quotecolor}{HTML}{000000}
\newcommand{\highlightquote}[1]{%
  {\textcolor{quotecolor}{\textit{``#1''}}}%
}

\definecolor{lkVU}{HTML}{6F7C80} 
\definecolor{lkU}{HTML}{8C979A}  
\definecolor{lkN}{HTML}{A7AFB2}  
\definecolor{lkL}{HTML}{E6A23C} 
\definecolor{lkVL}{HTML}{C47A00} 
\newcommand{\likertnumfont}{\small}  
\def\likertlabelmin{10} 

\newcommand{\likertCell}[6]{%
\begin{tikzpicture}[baseline=-0.5ex]
  \def\sc{0.028}%
  \pgfmathsetmacro{\xvl}{#5*\sc}%
  \pgfmathsetmacro{\xl}{\xvl+#4*\sc}%
  \pgfmathsetmacro{\xn}{\xl+#3*\sc}%
  \pgfmathsetmacro{\xu}{\xn+#2*\sc}%
  \pgfmathsetmacro{\xvu}{\xu+#1*\sc}%
  \useasboundingbox (0,-3.8ex) rectangle (2.8,1.45ex);
  \fill[lkVL] (0,-1.2ex)     rectangle (\xvl,1.2ex);
  \fill[lkL]  (\xvl,-1.2ex)  rectangle (\xl,1.2ex);
  \fill[lkN]  (\xl,-1.2ex)   rectangle (\xn,1.2ex);
  \fill[lkU]  (\xn,-1.2ex)   rectangle (\xu,1.2ex);
  \fill[lkVU] (\xu,-1.2ex)   rectangle (\xvu,1.2ex);
  \draw[black,line width=0.6pt] (\xl,-1.35ex) -- (\xl,1.35ex);
  \draw[black,line width=0.5pt] (0,-1.8ex) -- (\xl,-1.8ex);
  \pgfmathsetmacro{\mendx}{\xl/2}%
  \node[font=\likertnumfont, text=black] at (\mendx,-3.0ex) {#6\%};
  \pgfmathsetmacro{\mvl}{\xvl/2}
  \ifnum#5>\likertlabelmin\relax
    \node[font=\likertnumfont,text=black] at (\mvl,0){#5};
  \fi
  \pgfmathsetmacro{\ml}{(\xvl+\xl)/2}
  \ifnum#4>\likertlabelmin\relax
    \node[font=\likertnumfont,text=black] at (\ml,0){#4};
  \fi
  \pgfmathsetmacro{\mn}{(\xl+\xn)/2}
  \ifnum#3>\likertlabelmin\relax
    \node[font=\likertnumfont,text=black] at (\mn,0){#3};
  \fi
  \pgfmathsetmacro{\mu}{(\xn+\xu)/2}
  \ifnum#2>\likertlabelmin\relax
    \node[font=\likertnumfont,text=black] at (\mu,0){#2};
  \fi
  \pgfmathsetmacro{\mvu}{(\xu+\xvu)/2}
  \ifnum#1>\likertlabelmin\relax
    \node[font=\likertnumfont,text=black] at (\mvu,0){#1};
  \fi
\end{tikzpicture}%
}

\usepackage{tikz}
\newcommand{\circled}[1]{%
  \tikz[baseline=(char.base)]{
    \node[shape=circle, draw, inner sep=1pt, font=\small] (char) {#1};
  }
}

\newcommand{\spark}[1]{%
  \raisebox{-0.22\height}{%
    \includegraphics[width=2.5cm,height=1.2cm,keepaspectratio]{figure/#1.png}}}

\newcommand{\evidence}[1]{\newline{\scriptsize\itshape #1}}
\newcommand{\timeline}[1]{\textcolor{lkVL}{\textbf{#1}}}

\definecolor{adoptLowBg}{RGB}{232,245,233}   
\definecolor{adoptLowFg}{RGB}{27,94,32}      

\definecolor{adoptMidBg}{RGB}{129,199,132}   
\definecolor{adoptMidFg}{RGB}{0,77,64}       

\definecolor{adoptHighBg}{RGB}{46,125,50}    
\definecolor{adoptHighFg}{RGB}{255,255,255}  

\newcommand{\adoptHigh}{%
{\small\bfseries\colorbox{adoptHighBg}{\textcolor{adoptHighFg}{$\blacktriangle$~High}}}}

\newcommand{\adoptMid}{%
{\small\bfseries\colorbox{adoptMidBg}{\textcolor{adoptMidFg}{$\blacksquare$~Medium}}}}

\newcommand{\adoptLow}{%
{\small\bfseries\colorbox{adoptLowBg}{\textcolor{adoptLowFg}{$\triangledown$~Low}}}}

\newcommand{\aiddos}[0]{AI-DDoS }

\begin{document}

\title{``AI Slop is DDoSing Open Source'': Understanding the Impact of AI-Generated Contributions on Open Source Sustainability}



\author{
\IEEEauthorblockN{
Sadia Afroz$^{1}$ \hspace{0.8em}
Courtney Miller$^{2}$ \hspace{0.8em}
Tyler Menezes$^{3}$ \hspace{0.8em}
Edward Gilmour$^{1}$ \hspace{0.8em}
Anita Sarma$^{1}$ \hspace{0.8em}
Zixuan Feng$^{4}$
}

\IEEEauthorblockA{
$^{1}$Oregon State University, Corvallis, OR, USA (\{afrozs, gilmoued, anita.sarma\}@oregonstate.edu)\\
$^{2}$The George Washington University, Washington, DC, USA (courtney.miller@gwu.edu)\\
$^{3}$CodeDay, Seattle, WA, USA (tylermenezes@codeday.org)\\
$^{4}$Virginia Commonwealth University, Richmond, VA, USA (fengz3@vcu.edu)
}
}

\maketitle

\begin{abstract}
Open source software (OSS) communities are facing increasing pressure from Generative AI (GenAI) tools.  We call it AI-DDoS: a denial-of-service effect in which plausible but low-quality AI-generated contributions overwhelm OSS community capacity. Using a phenomenon-based mixed-methods approach, we first analyze practitioner accounts from Reddit, OSS mentor mailing lists, and blogs to identify six recurring themes and derive hypotheses. We then evaluate these hypotheses using Bayesian Structural Time Series analysis across 294 repositories with over 2 million pull requests and issues. Our results show that while PR volume increased in 2025, merge rates declined, with one-time contributors experiencing an 18.18\% drop in PR merge rates relative to the counterfactual. Finally, we identify 11 remediation strategies through practitioners' interviews and validate them with a survey of 229 OSS practitioners, grouping them into preservative, adaptive, and transformative orientations. Our findings show that AI-DDoS is not only a contribution-volume problem but a sustainability trap: communities often default to low-effort defensive strategies that protect short-term review capacity while making openness difficult to sustain.

\end{abstract}


\input{sections/1_intro}

\label{sec:intro}

\input{sections/2_related_work}
\label{sec:related}

\input{sections/3_RQ1}

\input{sections/4_RQ2}

\input{sections/6_Limitations}

\input{sections/5_Discussion}

\bibliographystyle{IEEEtran}
\bibliography{bib}

\end{document}

%% file: sections/1_intro.tex
\section{Introduction}

\begin{figure}[t]
    \centering
    \includegraphics[width=\linewidth]{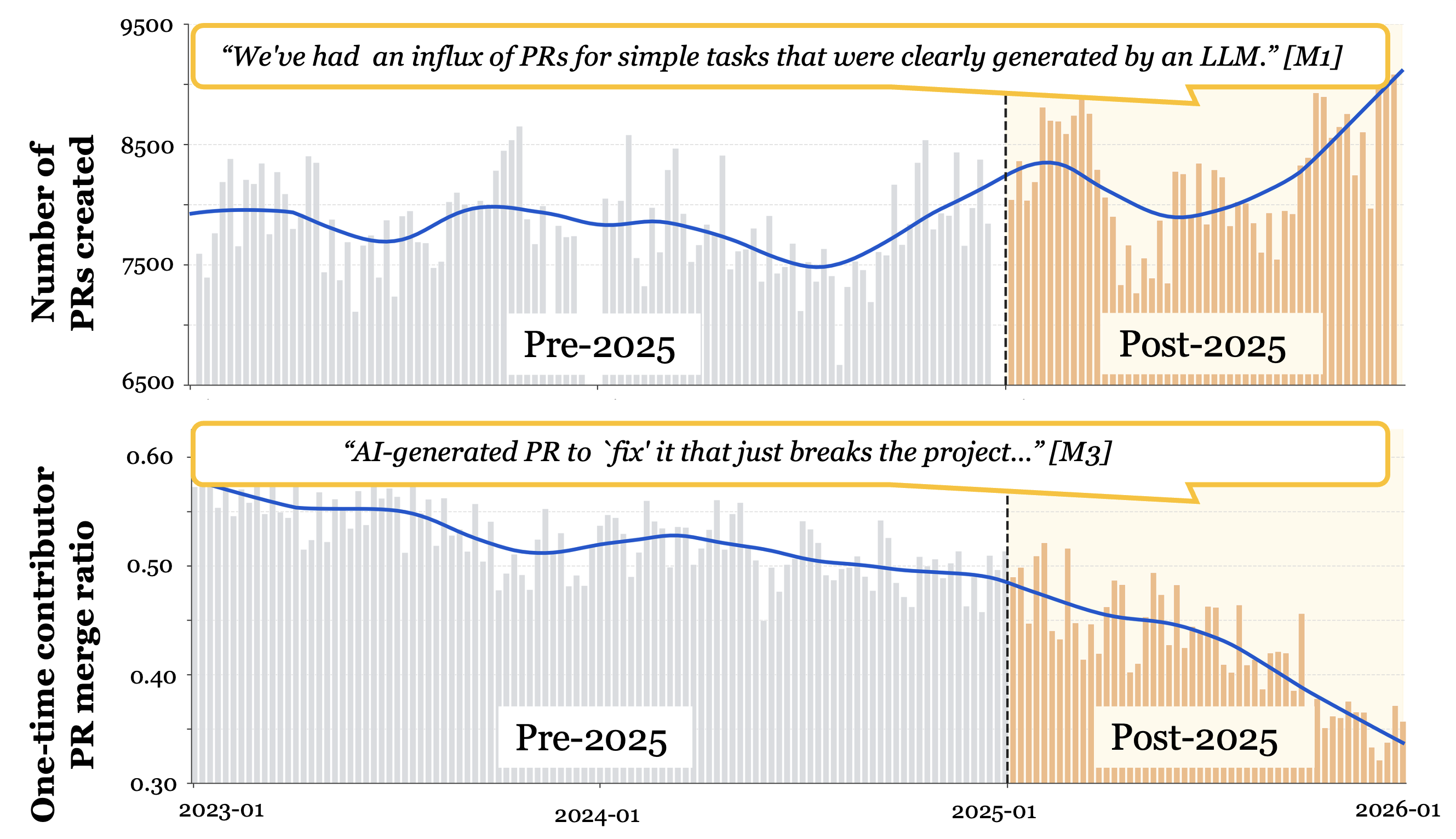}
    \vspace{-8pt}
    \caption{Overview of contribution flow.}
    \vspace{-8pt}
    \label{fig:overview}
\end{figure}

Generative Artificial Intelligence (GenAI) tools for software development have dramatically lowered the cost of producing code while raising the cost of evaluating it~\cite{afroz2026fast, GitHub2026}. In open source software (OSS) communities, this asymmetry produces a phenomenon we call \textit{AI-DDoS}: a flood of plausible-looking, low-quality GenAI contributions that disrupts the contribution process and overwhelms the limited time and resources maintainers have to review contributions. Figure \ref{fig:overview} presents an overview of the steep rise in contributions and the decline in accepted contributions.

\boldification{OSS projects are defensively responding to it such as cURL project ended its bug bounty program}
OSS practitioners have begun responding defensively to this barrage~\cite{Ghostty, claburn2025curl}. In January 2026, the cURL project ended its six-year bug bounty program after AI-generated contributions reduced the proportion of valid submissions to about 5\% by July 2025, making triage unsustainable \cite{claburn2025curl, itsfoss2026curl}. Others have moved to invitation-only contribution workflows, adopted blanket policies rejecting unsolicited pull requests, and stopped accepting outside vulnerability reports \cite{Ghostty, GitHub2026}.

\boldification{Such defensive responses may risk closing the open-source door}
\boldification{Such defensive responses risk closing the open-source door}
These responses help projects survive the immediate flood, but they narrow or completely block the pathways through which OSS sustains itself. Over a decade of OSS sustainability research establishes that a project's health and survival depend on its maintainers, who provide the labor that keeps the project alive, and on its ability to recruit and retain new contributors who renew that labor over time~\cite{guizani2022attracting, linaaker2024sustaining}. The problem is that the programs OSS projects are shuttering to cope with \aiddos are the same ones they use to recruit and retain those contributors (e.g., good first issues~\cite{goodfirstissueGithub}). \aiddos and the strategies used to remediate it therefore reach well beyond their immediate toll on the maintainers (e.g., Linux~\cite{james2026torvalds} and Ghostty~\cite{joshi2026ghostty}). The first impact is on the projects themselves: as communities close their doors to survive the flood, they cut off the contributor pipeline that renews maintainer labor, turning open projects into closed systems that slowly atrophy through attrition. The second impact is systemic: as these projects weaken, so do the critical software supply chains that depend on them~\cite{boughton2024decomposing}.

\boldification{However, no empirical study has systematically investigated the \aiddos phenomenon}
However, despite a rapidly growing body of practitioner accounts describing their lived experience with \aiddos~\cite{mathis2026aislop, james2026torvalds}, we lack an empirically grounded understanding of how prevalent it is and how it manifests across ecosystems. We also do not know whether \aiddos is a transient issue that existing contribution workflows can absorb, or a structural shift that the  contribution workflows cannot withstand without redesign.

In this paper, we address this gap through a phenomenon-based, mixed-methods study~\cite{creswell2017designing}. Phenomenon-based research (\textit{PBR})~\cite{lumineau2025roadmap} aims to identify, document, and conceptualize an emerging phenomenon to advance knowledge through phenomenon selection, framing with stakeholders or theoretical grounding, and data collection. Following this approach, we characterize the phenomenon \aiddos as an emerging form of structural pressure on OSS, examine how it affects contribution workflows in practice, and identify remediation strategies for maintainers and the broader OSS community. We ask two research questions:
\begin{itemize}
    \item \textbf{RQ1.} \textit{How does \aiddos manifest in OSS contribution workflows?}
    \item \textbf{RQ2.} \textit{What \aiddos remediation strategies can maintainers and the broader OSS community adopt?}
\end{itemize}

To answer \textbf{RQ1}, we first triangulate practitioner accounts across three independent sources of gray literature (Reddit, practitioner blogs and the OSS mentors' mailing list) to characterize how AI-DDoS manifests in practice, identify the practitioner-reported mechanisms underlying these manifestations, and derive hypotheses about how it would appear in repository-level contribution data. We then test these hypotheses at ecosystem scale, using Bayesian Structural Time Series (BSTS) causal impact analysis across 294 widely-used projects spanning more than 2 Million pull requests and issues. We treat 2025 as the intervention boundary, the year in which practitioner reports converged and AI-assisted contribution became a routine part of OSS workflows~\cite{StackOverflow2025survey, github2025octoverse}. This study design lets us move from what practitioners report experiencing to whether those signals appear in the contribution workflows of real projects.

To answer \textbf{RQ2}, we conducted semi-structured interviews with 11 OSS maintainers, guided by Organizational Resilience Theory~\cite{vogus2007organizational, hillmann2021organizational}, to identify the strategies communities are mobilizing against \aiddos. We then surveyed 229 OSS practitioners to test how widely these strategies are accepted across the OSS ecosystem.


Across both RQs, the evidence converges: \aiddos is an emerging, ecosystem-wide structural crisis, not an isolated complaint. Practitioner accounts and repository data agree on a post-2025 shift in which contribution volume rises while quality and completion decline. 

In summary, we contribute: \textbf{(1)} A characterization of how the \aiddos phenomenon manifests in practice, drawn from practitioner-reported themes triangulated across three independent sources of gray literature.
\textbf{(2)} Causal impact evidence that \aiddos is a structural, ecosystem-wide problem: across 294 widely-used projects, PR volume rose while merge rates and issue completion fell after 2025, with the sharpest effect among one-time contributors. 
\textbf{(3)} A set of 11 practitioner-driven, validated  (survey with N=229) remediation strategies that surfaces a central tension: communities favor low-effort preservative strategies that can harden into defensive closure over time.

%% file: sections/2_related_work.tex
\section{Background}
\label{sec:background}

\textbf{Open participation is a long-term investment.}
OSS sustainability is a project's ability to continuously generate the labor required to maintain it, through sustained maintainer engagement, the attraction and retention of contributors, and the conversion of newcomers into core developers~\cite{guizani2022attracting, miller2019people}. The well researched ``\emph{onion model}'' describes how contributors move from passive users to active users, patch contributors, and eventually core developers~\cite{ye2003toward, crowston2005, jergensen2011onion}. 

Prior work has therefore examined how OSS projects can keep this pathway open by lowering entry barriers through ``good first issues'', mentoring programs, contribution guidelines, and structured onboarding practices~\cite{steinmacher2019overcoming, feng2024guiding}. Yet this model of sustainability relies on trust: trust that outsiders arrive with good-faith intent, that maintainers can distinguish promising newcomers from low-quality noise, and that welcoming many outsiders will replenish the contributor base~\cite{zhou2016inflow}. When that trust fails, openness becomes a liability.

\textbf{AI-assisted tools weaken this trust.}
For example, reviewing a contribution requires maintainers to assess not only functional correctness, but also architectural fit, test coverage, and long-term maintainability~\cite{kalliamvakou2014promises, steinmacher2018almost}. In the past, this review burden was partly constrained by the effort required to produce a plausible contribution, which demanded project knowledge and sustained effort~\cite{tidelift2024maintainer}. Today, AI-assisted tools weaken that constraint by making plausible-looking contributions faster and cheaper to produce~\cite{pimenova2025good, leiter2024chatgpt}, while the labor needed to maintain them remains limited and exhaustible~\cite{feng2025gains, zhou2016inflow}.

The 2025 Stack Overflow Developer Survey reports that 84\% of developers use or plan to use them and 51\% use them daily, while GitHub’s Octoverse 2025 reports platform-scale growth in developers and merged pull requests~\cite{StackOverflow2025survey, github2025octoverse}. Controlled studies report productivity gains\cite{brynjolfsson2025generative, DellAcqua2023}, especially for less experienced developers~\cite{peng2023impact}, but emerging evidence shows quality tradeoffs, including increased churn, reduced reuse, more static-analysis warnings, and higher complexity~\cite{GitClear2025, he2025speed}. GitHub describes 2025 as an OSS ``Eternal September'', where rising contribution volume overwhelms human attention and strains trust~\cite{GitHub2026}. CodeRabbit reports that AI-generated pull requests contain 1.7 times more issues than human-written ones~\cite{CodeRabbit2025}.

\textbf{OSS projects are putting up additional gates.}
The impact of the AI-assisted contribution pressure is already visible at the level of individual projects, and the responses it provokes reveal the deeper sustainability threat. In January 2026, cURL ended its six-year bug bounty program after AI-generated reports pushed the valid-submission rate from roughly one in six to one in twenty or thirty ~\cite{stenberg2026fosdem}. Other projects have responded by adding stronger gates around outside contributions: Ghostty requires disclosure of AI use, human understanding of submitted code, and sanctions for low-effort AI-generated work~\cite{Ghostty}. The Linux kernel has similarly responded to duplicate AI-generated vulnerability reports by redirecting them away from the private security list, while requiring AI-assisted contributions to remain transparent and legally accountable through human sign-off and the \texttt{Assisted-by} tag \cite{james2026torvalds}.

These responses may help projects survive the immediate pressure, but at the cost of narrowing the open pathways through which OSS sustains. \emph{This creates an urgent need to investigate GenAI-driven contribution pressure in OSS: how it reshapes OSS sustainability, and how communities respond to openness becoming a source of strain.}

%% file: sections/3_RQ1.tex
\section{The Genesis of AI-DDoS (RQ1)}
\label{sec:rq1}
To answer \textbf{RQ1} through two stages, following the PBR guidance on phenomenon selection and data collection for emerging phenomena~\cite{lumineau2025roadmap}: Stage 1 characterizes \aiddos phenomenon by triangulating practitioner-reported mechanisms across three sources of gray literature, and Stage 2 validates whether the phenomenon appears in repository-level contribution traces using causal impact analysis.

\subsection{Stage 1 Method: Triangulated Qualitative Analysis}
\label{sec:rq1_stage1}

We begin by analyzing practitioner reports from gray literature to characterize AI-DDoS as an emerging practitioner-reported phenomenon and to develop hypotheses for quantitative validation. Because AI-DDoS is recent and still unfolding in OSS, gray literature provides timely evidence of early signals, community concerns, and practitioner responses, and has been widely used in empirical SE studies~\cite{garousi2019gray, kitchenham2009systematic}.

\subsubsection{Data Collection} To identify relevant artifacts, we use keyword-based search for \emph{practitioner blogs} and  \emph{Reddit (r/opensource)} followed by an exhaustive review of the discussions on the \emph{mentors' mailing list}, following established SE practices for analyzing gray literature~\cite{garousi2019gray}. These three independent sources enable data triangulation, strengthening the credibility of our qualitative findings.

\emph{Practitioner Blogs.} To determine which practitioner blogs to include in our search, we followed standard blog inclusion approaches used in prior SE studies~\cite{garousi2019gray, fengaddressing}. We first identified the top 50 OSS blog platforms from Feedly’s ``Best Open Source Blogs and Websites'' list (\url{https://feedly.com}; accessed March 2026). We excluded project-specific and primarily release-announcement-driven blogs or those lacking editorial curation, resulting in 32 eligible platforms. These included foundation- and community-governed platforms such as the Open Source Initiative~\cite{osi_blog_2026}, Google Open Source Blog~\cite{google_open_source_blog_2026}, and Linux.com~\cite{linux_com_2026}.

\boldification{To collect the blogs, we first used two blogs to generate search keywords; our search yielded 29 blogs}
To retrieve relevant blogs, we first developed an initial keyword set from two blog posts \cite{holterhoff2026slopageddon, iyer2026codecrisis} already known to the research team. These posts discussed AI-generated/-assisted contributions and provided seed terminology for the search process. From this initial review, we identified nine search terms \cite{supply}, such as \textit{AI-generated pull requests} and \textit{low-quality contributions}. We used those to retrieve potentially relevant blogs. Three researchers independently reviewed each candidate blog artifact for relevance. They then met weekly to discuss disagreements and reach an agreement on the final blog set. This resulted in 12 relevant blog post artifacts (B1--B12). See the supplementary materials for the bloglists \cite{supply}.

\boldification{Reddit captures reactive, time-sensitive discourse.}
\emph{Reddit OSS community \textit{r/opensource}} captures reactive, time-sensitive discourse from a broad community of practice~\cite{wenger2000communities} reflecting immediate experiences, concerns, and observations \cite{bergstrom2022signaling}. Prior works have shown that such developer-facing social platforms can surface early indicators of socio-technical change and evolving practitioner concerns~\cite{pimenova2025good, newman2025get}. We apply the same set of keywords used to search for relevant blog posts to identify relevant discussions within this subreddit community (\textit{r/opensource}), yielding 34 candidate threads. Three researchers independently assessed each thread for relevance and reached full agreement through weekly negotiated agreement sessions. During these sessions, ten threads were excluded because they were not specific to OSS, resulting in a final corpus of 24 threads (R1-R24) comprising 334 comments.

\boldification{MML provides institutionally mediated signals.}
\emph{OSS Mentors Mailing List (MML)} captures practitioner communication in a structured, multi-organization OSS environment, where mentors coordinate expectations, practices, and responses across participating OSS organizations. Recurring discussions in the mailing list provide evidence of emerging ecosystem-level concerns~\cite{ steinmacher2013newcomers}.

We obtained access to the full mentors’ mailing-list archive in accordance with IRB-approved procedures. Given our focus on AI-assisted contributions, we reviewed five years of discussion history (2022–2026) and conducted an exhaustive review of all email discussions. Three researchers independently assessed all emails for relevance, with discrepancies resolved through weekly meetings. This process identified 13 relevant discussion threads (M1–M13), comprising 119 emails.

\subsubsection{Qualitative Analysis.} 
\boldification{We conducted thematic analysis...}
We used reflexive thematic analysis~\cite{braun2006using} for all artifacts to characterize community reactions, practitioner experiences, and emerging pressures associated with AI-generated/-assisted contributions. Three authors iteratively analyzed the dataset, developing preliminary codes and discussing emerging interpretations in weekly meetings. Through these discussions, we refined our understanding of the data, reorganized related codes, and developed higher-level themes, which informed our characterization of AI-DDoS as a form of contribution pressure on OSS communities.

\input{Tables/theme_dist}

\subsection{Stage 1 Results: AI-DDoS as a Progressive OSS Breakdown}
\label{sec:rq1_stage1_results}

Our qualitative analysis reveals \textbf{AI-DDoS} as an emerging practitioner-reported phenomenon in OSS: a denial-of-service \cite{garber2000denial} effect in which AI-generated contributions overwhelm a project's capacity to review, triage, validate, and absorb contributions. In the following sections, we unpack this phenomenon through six recurring themes, as shown in Table~\ref{tab:themes_dist}. These themes show cross-source convergence among the three data sets we analyzed, with a concentration in 2025.

\textbf{T1: AI-DDoS begins as a traffic flood.} \highlightquote{AI slop flooding open source}~[B3] \cite{mathis2026aislop}, and projects having \highlightquote{...bug reports with AI-generated contents that don't offer any real value}~[B9] \cite{rudra2025curl_ai_slop}. \highlightquote{We've had an influx of PRs for simple tasks that were clearly generated by an LLM.}~[M1]. In bug bounty and security-reporting systems, the projects were \highlightquote{inundated with AI slop, where many bogus reports were opened}~[B10] \cite{itsfoss2026curl} as \highlightquote{Now they face script kiddies opening Cursor, finding bugs, generating fixes, and submitting pull requests without understanding the code.}~[B3] \cite{mathis2026aislop}. The speed of the flood was also visible in the cases where \highlightquote{seven [X] reports came in within a 16-hour period.}~[B10] \cite{itsfoss2026curl}.


\textbf{T2: The flood consists of low-quality contributions.} Practitioners described this high traffic as \highlightquote{PRs that compile but are unreviewable}[R1], and \highlightquote{...low-quality submissions where it isn't obvious whether the submission came from a person or an AI model}[B1] \cite{claburn2025curl}. Some submissions contained fabricated evidence, such as \highlightquote{AI-generated `screenshots' of non-existent [X] UI}~[M4]. In more harmful cases, AI-generated work appeared as a full issue-to-PR chain: \highlightquote{opening an AI-generated nonsense issue and then an AI-generated PR to `fix' it that just breaks the project by deleting large blocks of code to improve `performance'}~[M2].


This leads us to \textbf{Phenomenon 1 (Low quality traffic overload)}: AI-DDoS begins when OSS projects receive a high volume of AI-generated contributions that are often low-quality, fabricated, or harmful.

\textbf{T3: Drive-by contributions.} Practitioners linked AI-DDoS to transient contributors who contribute once and never engage with the community again, using AI to generate drive-by contributions. Some projects have already begun taking steps to address this pattern \cite{rudra2025gnome}. \highlightquote{it's low-effort drive-by contributions from people who don't understand the codebase. AI just made it cheaper to produce them at scale}~[R3]. \highlightquote{We can't do anything about drive-by submissions from those who don't engage with the org.}~[M3].

\textbf{T4: Drive-by contributions are amplified by incentive abuse.} Practitioners connected these drive-by contributions to external incentives around GitHub visibility and the job market \cite{itsfoss2026curl, rudra2026vouch}. \highlightquote{I have definitely noticed an increase in people trying to exploit open source. Markets are tough, so there's incentive to farm GitHub contributions for resumes}~[R14]. \highlightquote{The system right now incentivizes: push code fast, get PRs merged, pad the resume. AI just lowered the barrier to generating volume}~[R1]. Practitioners warned that \highlightquote{Some people are trying to build up realistic looking GitHub profiles so that they can do supply line attacks}~[R8].

These accounts point to \textbf{Phenomenon 2 (Incentivized drive-By contribution)}: AI lowers the cost of generating visible OSS contribution traces, amplifying profile-building behavior beyond one-time contributors.

\textbf{T5: Denial-of-service effect: Human Breakdown.} The first breakdown produced by AI-DDoS occurs at the human layer \cite{mathis2026aislop, claburn2025curl}. Practitioners reported that \highlightquote{it's just not realistic for any maintainer to look at the sheer volume of incoming pull requests that we're getting right now} [M9]. \highlightquote{The deluge of PRs is just wasting our time}~[M7]. Maintainers must now go through \highlightquote{the torrent of nonsensical PRs generated by LLMs, in search of the handful of PRs by genuine new contributors}~[R1], and even \highlightquote{rejecting them takes more time than just writing the code}~[B12] \cite{rudra2026vouch}.

The breakdown is not only a matter of workload. Before AI-generated submissions, mistakes in pull requests could signal genuine learning; reviewers could identify where contributors were confused and help them improve~\cite{feng2022case}. \highlightquote{I've enjoyed reviewing PRs in the past, often with mistakes as many make their first PR. Because we know that that is a genuine effort to learn. But now it feels like we are doing free work for robots to validate their code}~[M9]. \highlightquote{And now am I supposed to bond with Claude or Gemini? In the past, I can clearly see from the patch our contributors submitted what confusion they had, and the tough design decisions they had to make... Now these gaps are smeared by LLM}~[M10].

\textbf{T6: Denial-of-service effect: System Breakdown.} As human capacity becomes overwhelmed, AI-DDoS begins to reshape the openness. \highlightquote{We've had to stop assigning issues at all}~[M7]. Some practitioners described more formal exclusionary responses: \highlightquote{It's so bad that this is the first year in our 10 years of existence that we have ever blocked contributors}~[M6]. In some cases, the pressure led not only to gating but also to withdrawal, \highlightquote{Python has multiple sub orgs taking the year off due to burnout, so you're not alone}~[M7].

We therefore characterize \textbf{Phenomenon 3}: (Denial-of-service effect): AI-DDoS first breaks the human layer of OSS; communities then protect themselves by narrowing the open channels that once sustained broad participation.

\subsection{Stage 2 Method. Quantitative Validation with Repo Data}

\boldification{Stage 1 only identified practitioner-reported signals, but maybe biased...}
Stage 1 characterizes practitioner-reported AI-DDoS, with evidence converging around 2025 across three independent sources. Stage 2 examines whether this phenomenon corresponds to repository-level changes.

Guided by the themes from Stage 1 (Table \ref{tab:themes_dist}), we developed four primary hypotheses on contribution volume and quality, and then extended them to contributor-behavior hypotheses. The hypothesis-theme mapping is provided in the supplementary material \cite{supply}. We then use a Bayesian Structural Time Series (BSTS) causal impact framework \cite{brodersen2017causalimpact} to test whether repository-level behaviors changed significantly after widespread AI-assisted contribution emerged. BSTS is well suited to our study because our objective is to estimate the counterfactual trajectory of repository activity following an ecosystem-wide intervention for which no natural comparison group exists. It learns the pre-intervention temporal dynamics and projects the expected post-intervention trajectory under the no-intervention scenario which allows us to compare the observed outcomes with their estimated counterfactual.


\textbf{Themes to Hypotheses.} As shown in Table~\ref{tab:themes_dist} (T1, T2, T4, and T5), practitioners across all three sources reported that AI-assisted contributions \textbf{increase contribution volume while reducing contribution quality}. Therefore, we propose four primary hypotheses: \textbf{H1: After AI adoption, contribution activity changes as follows}: PR volume increases (\textbf{H1a}); PR merge ratio decreases (\textbf{H1b}); issue volume increases (\textbf{H1c}); and issue completion ratio decreases (\textbf{H1d}).

T3 concerns \emph{one-time contributors} (contributors who submitted exactly one PR or issue to a repository). Yet the incentive and platform structures associated with AI-assisted contribution (T4) may reshape participation among both transient (one-time) and recurring (\emph{non-one-time}) contributors. To examine whether aggregate shifts are driven by different contributor groups, we extend H1 by decomposing contribution activity into recurring contributors and transient contributors. Accordingly, we hypothesize: \textbf{H2: Among non-one-time contributors, the H1 patterns are reflected after AI adoption}: PR volume increases (\textbf{H2a}); PR merge ratio decreases (\textbf{H2b}); issue volume increases (\textbf{H2c}); and issue completion ratio decreases (\textbf{H2d}). \textbf{H3: Among one-time contributors, the H1 patterns are reflected after AI adoption}: PR volume increases (\textbf{H3a}); PR merge ratio decreases (\textbf{H3b}); issue volume increases (\textbf{H3c}); and issue completion ratio decreases (\textbf{H3d}).

\subsubsection{From Qualitative Signals to a Repository Sample}

To prepare for the analysis, we scoped our repo sampling pool using five criteria. (i) We started from a curated directory of actively maintained repositories (\url{https://goodfirstissue.dev/}; accessed March 2026) to focus on OSS projects with observable external contribution activity. (ii) We retained repositories with at least 5{,}000 GitHub stars to identify widely adopted OSS projects with active contributor communities, following prior OSS studies~\cite{steinmacher2018almost, de2024funless}, yielding 355 candidates. (iii) To ensure sustained contribution traffic, we retained repositories with at least 100 contributions per year and excluded repositories with inactivity gaps longer than six months~\cite{kalliamvakou2014promises}. (iv) We retained repositories created no later than 2023 and continuously active through 2025 to compare contribution activities before and after AI adoption~\cite{stackoverflow2023survey, stackoverflow2024survey, StackOverflow2025survey}. (v) Finally, we removed archived repositories and those with disabled issues or pull requests to ensure access to contribution-related interactions.

After applying these criteria, we retained 294 repositories, including foundation-governed projects (e.g., \textit{Kubernetes}, \textit{Elasticsearch}), industry-backed platforms (e.g., \textit{JetBrains}, \textit{Alibaba}), and widely used developer tools (e.g., \textit{Grafana}). Then, using the GitHub REST API, we collected PR and issue activity from January 2023 to December 2025, including creation timestamps, contributor identifiers, and outcome status. Because our analysis focuses on human participation, we removed automated accounts with bot markers \texttt{[bot]}, \texttt{-bot}, or \texttt{\_bot}. The \textbf{final dataset contains 1,224,374 PRs and 822,133 issues from 356,972 unique contributors.}

\subsubsection{Data Preparation}

Before conducting the analysis, we perform two preprocessing steps: (i) \textit{Temporal granularity: weekly.} We aggregate PRs and issues by week to balance temporal sensitivity and model stability for BSTS, as daily data are often sparse and noisy and  monthly data may obscure shifts that unfold over shorter periods.(ii) \textit{Aggregation: 
volume-weighted across repositories.} Because repositories vary substantially in contribution volume, we aggregate weekly metrics across repositories using weights proportional to each repository's weekly contribution volume to better reflect ecosystem-level activity~\cite{chikezie2025measuring, feng2025gains}.


\subsubsection{Causal Impact Analysis via Bayesian Structural Time Series (BSTS)}

To assess whether AI-assisted contributions changed OSS workflows, we estimate counterfactual trajectories for each outcome metric using a univariate BSTS model implemented through the \texttt{CausalImpact} package~\cite{ brodersen2017causalimpact}. The model learns pre-intervention level without covariates and projects them into the post-intervention period to estimate the expected trajectory had pre-intervention trends continued, following prior studies~\cite{brodersen2015inferring, thorakkattle2022forecasting}.

We treat 2025 as the intervention period as Stage 1 identified consistent practitioner discussions across all three data sources during 2025 (Table~\ref{tab:themes_dist}), indicating that AI-assisted contribution had become a visible OSS workflow phenomenon, consistent with recent tech reports~\cite{StackOverflow2025survey, github2025octoverse}. Given the gradual diffusion of AI-adoption across OSS ecosystems, we use a calendar-year boundary instead of a specific date to avoid false precision while providing a pre-specified intervention point for causal impact analysis~\cite{wagner2002segmented, bernal2017interrupted}. Therefore, in our setting, BSTS learns each metric's expected level and variation from January 2023--December 2024 and predicts this baseline into 2025 to estimate how the metric would have evolved without the surge of AI-assisted contribution in 2025.

We first assessed model robustness using placebo tests~\cite{lenhart2021effects}, setting January 2024 as a pseudo-intervention date for the H1 outcomes. The tests showed no significant placebo effects for the PR-related metrics, but revealed significant placebo effects for issue-related metrics, which we discuss in Section~\ref{sec:threats}.

Table~\ref{tab:causal-results} reports the BSTS results for each hypothesis. The \textit{Abs.} column represents the average weekly deviation from the counterfactual and the \textit{Rel.} column reports the corresponding percentage change from the expected trajectory~\cite{brodersen2015inferring}. Posterior probability indicates whether the true effect falls within the hypothesized interval and serves as our primary support criterion, using a 95\% threshold~\cite{brodersen2015inferring}. Cohen's $d$ standardizes the effect by pre-intervention variability~\cite{cohen2013statistical}.  We also visualize weekly temporal trends using LOESS smoothing~\cite{basak2023challenges}, with a dashed vertical line marking the January 2025 AI-surge intervention.

\input{Tables/causal_results}

\subsection{Stage 2 Results: Quantitative Validation of AI-DDoS}
\label{sec:stage_2_results}



\textbf{H1a. Increased PR Influx} is supported with a medium effect size. As shown in Table~\ref{tab:causal-results}, the temporal plot shows a visible increase in PR volume after AI adoption. The BSTS estimates indicate that weekly PR volume ran $6.80$\% above the counterfactual during 2025. The \emph{absolute} burden is substantial: the effect corresponds to roughly $520$ excess PRs per week across the 294 sampled repositories, accumulating to approximately $27{,}000$ additional PRs over the one-year post-intervention window. An illustrative case is \textit{elastic/kibana}. In that sample, the average weekly PR volume increased from $303$ in 2023 to $364$ in 2024, a $20$\%. This moderate year-over-year increase contrasts with the substantially larger $50$\% increase observed in 2025, when average weekly PR volume reached $545$. In absolute terms, this shifted the repository from roughly $19{,}000$ PRs in 2024 to about $28{,}000$ PRs in 2025.

\textbf{H1b. Reduced PR Merge Ratio} is supported with a medium effect size. After AI adoption, the temporal plot shows a downward shift in the PR merge ratio. The BSTS estimates indicate that the size-weighted merge ratio was about $0.01$ below the counterfactual on average weekly, corresponding to a $1.06$\% reduction. When standardized against the metric's own pre-intervention stability, this shift reaches a medium effect size ($|d|= 0.68$, posterior probability $99.9$\%). This downward displacement suggests reduced mergeability in 2025. The increased PR influx (H1a) and declining mergeability (H1b) together provide quantitative evidence for Phenomenon~1 (low quality traffic overload).

\textbf{H1c. Issue Volume} is not supported but is significant in the opposite direction: issue volume ran $8.12$\% below the counterfactual, corresponding to an average of about $431$ fewer issues per week (roughly $22{,}000$ fewer over the post-intervention year). \textbf{H1d. Issue Completion Ratio}, however, is supported. The completion ratio was approximately $0.03$ below the counterfactual, a $3.40$\% relative reduction with a large effect size ($|d|=1.57$), indicating a substantial shift relative to its typical week-to-week variability.

This does not imply that AI-assisted contributions had no effect on issue workflows. The temporal pattern in Table~\ref{tab:causal-results} shows that issue volume declines during the early post-intervention period but rebounds sharply toward the end of the observation window, as highlighted in orange. Importantly, this rebound coincides with the steepest decline in the issue completion ratio, also highlighted in orange. We interpret this as the delayed propagation of AI's impact on the issue side.  Coding-assistant tools matured earlier in 2025, which likely shaped the PR-side behavior first, while agentic tools that scan codebases and generate bug reports are emerging now~\cite{agent2025fixbugs}.

\textbf{H2. The influx propagates to non-one-time contributors.} The results for \emph{recurring contributors} follow the same aggregate pattern observed in H1. \textbf{H2a} is supported: non-one-time contributors submitted $6.91$\% more PRs per week than expected ($d = 0.69$). This increase was accompanied by reduced PR merge ratio. \textbf{H2b} is also supported, with their PR merge ratio falling $3.45$\% below the counterfactual ($d = 2.14$).

On the issue side, recurring contributors exhibit a similar pattern as H1. \textbf{H2c} is not supported, as issue volume decreased by $12.40$\% weekly on average under counterfactual($d = 0.97$) , while \textbf{H2d} is supported, with the weekly issue completion ratio decreasing by $8.09$\% ($d = 1.60$).

\textbf{H3. One-time contributor burden.}
As shown in Table~\ref{tab:causal-results}, one-time contributors submitted more PRs after AI adoption, but these PRs were much less likely to be merged. \textbf{H3a} is supported: weekly one-time PR volume increased by $5.84$\% relative to the counterfactual ($d = 0.43$). \textbf{H3b} is also supported and shows the strongest effect in our analysis: the one-time PR merge ratio was $18.18$\% lower than expected ($d = 2.84$).

On the issue side, \textbf{H3c} is not supported, but \textbf{H3d} is: the weekly one-time issue completion ratio was, on average, $10.30$\% lower than expected ($d = 1.22$). The temporal plot shows that one-time issue volume initially declined but rebounded sharply late in the observation window, aligning with the aggregate issue pattern and suggesting that issue-side impacts emerged later. Practitioners described a similar PR-first dynamic, in which contributors were \highlightquote{gobbling up}~[M7] issues, especially \highlightquote{monopolizing the simplest 'good first issues' in the repo}~[M7].


%% file: Tables/theme_dist.tex
\begin{table}[t]
\centering
\caption{Qualitative themes with segment counts in sources.}
\vspace{-4pt}
\label{tab:themes_dist}
\setlength{\tabcolsep}{1.3pt}
\renewcommand{\arraystretch}{0.98}
\makebox[\columnwidth][c]{%
\begin{tabularx}{\columnwidth}{
>{\raggedright\arraybackslash}p{0.38\columnwidth}
>{\raggedright\arraybackslash}p{0.4\columnwidth}
>{\centering\arraybackslash}p{0.038\columnwidth}
>{\centering\arraybackslash}p{0.038\columnwidth}
>{\centering\arraybackslash}p{0.038\columnwidth}
>{\centering\arraybackslash}p{0.052\columnwidth}
}
\toprule
\textbf{Theme} & \textbf{Description}
&  \textbf{R}
& \textbf{B}
&  \textbf{M}
& \textbf{Tot.} \\
\midrule
\textbf{T1. Volume and submission flood}
\evidence{Blog \timeline{May'25}; Reddit \timeline{Nov'23}; MML \timeline{Mar'25}}
&
AI tools generate contributions (e.g., PRs, issues) at a rate that exceeds review capacity.
&
  16 &   20 &   41 & \textbf{77} \\
\rowcolor{gray!8}
\textbf{T2. Quality collapse of submissions}
\evidence{Blog \timeline{May'25}; Reddit \timeline{Apr'23}; MML \timeline{Mar'25}}
&
AI submissions look plausible but fail to meet quality expectations.
&
  11 &   7 &   17 & \textbf{35} \\
\textbf{T3. Transient contributor}
\evidence{Blog \timeline{May'25}; Reddit \timeline{Sep'25}; MML \timeline{Nov'25}}
&
Contributors use AI to mass-generate drive-by submissions.
&
  8 &   4 &   10 & \textbf{22} \\
\rowcolor{gray!8}
\textbf{T4. Incentive and platform structure}
\evidence{Blog \timeline{May'25}; Reddit \timeline{Sep'25}; MML \timeline{Mar'25}}
&
Structural incentives such as bounties, heatmaps, and pay-to-play stratification magnify AI-assisted submission.
&
  9 &   9 &   1 & \textbf{19} \\
\textbf{T5. Maintainer and mentor wellbeing}
\evidence{Blog \timeline{May'25}; Reddit \timeline{Jan'26}; MML \timeline{Mar'26}}
&
The flood overexhausts mentors and maintainers, driving burnout and withdrawal.
&
  4 &   6 &   9 & \textbf{19} \\
\rowcolor{gray!8}
\textbf{T6. Defensive contraction}
\evidence{Blog \timeline{May'25}; Reddit \timeline{Aug'25}; MML \timeline{Apr'25}}
&
Communities respond by closing channels, blocking by default, and restricting newcomer access.
&
  7 &  15 &   11 & \textbf{33} \\
\bottomrule
\end{tabularx}%
}
\begin{minipage}{0.98\columnwidth}
\scriptsize
 R = Reddit, B = Practitioner blogs, and M = Mentors mailing list discussions
\end{minipage}
\vspace{-10pt}
\end{table}

%% file: Tables/causal_results.tex
\begin{table}[t]
\centering
\caption{BSTS analysis results.}
\vspace{-4pt}
\label{tab:causal-results}

\scriptsize

\normalsize
\resizebox{\columnwidth}{!}{%
\fontsize{14}{16}\selectfont
\begin{tabular}{
@{}c l r r c c>{\centering\arraybackslash}m{2.5cm}@{}}
\toprule
\textbf{Hyp.} & \textbf{Metric} &
\textbf{Abs./Weekly} & \textbf{Rel.} &
\textbf{Post.} & \textbf{ $\lvert d\rvert$} & \textbf{Trend} \\
\midrule

\rowcolor{gray!12}\multicolumn{7}{@{}l}{\textit{Pull Requests}}\\
H1.a & \textbf{PR volume$^\ast$}      & 519.86  & +6.80\%  & 99.9\% & 0.69 & \spark{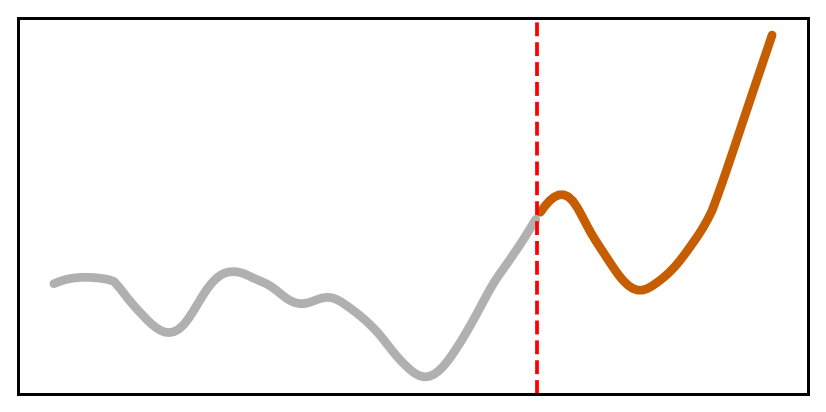}\\
H1.b & \textbf{PR merge.$^\ast$}     & -0.01  & -1.06\%  & 99.9\% & 0.68 & \spark{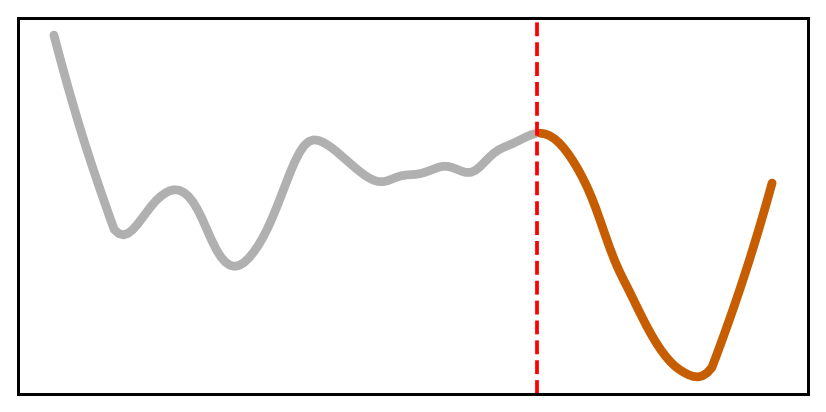}\\

\addlinespace[1pt]
\rowcolor{gray!12}\multicolumn{7}{@{}l}{\textit{Issue Requests}}\\
H1c & Issue volume$^\ast$   & -431.15 & -8.12\%  & 99.1\% & 0.65 & \spark{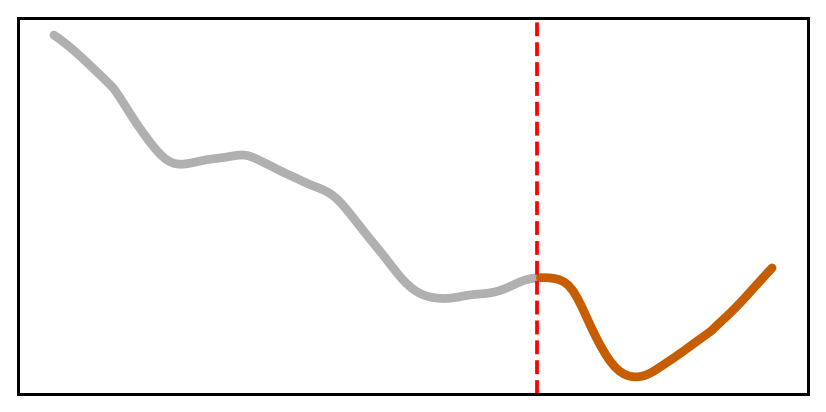}\\
H1d & \textbf{Issue compl.$^\ast$} & -0.03  & -3.40\%  & 99.1\% & 1.57 & \spark{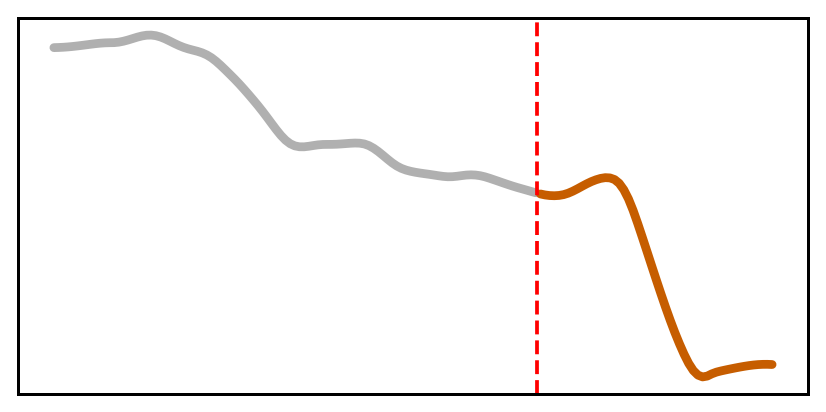}\\

\addlinespace[1pt]
\rowcolor{gray!12}\multicolumn{7}{@{}l}{\textit{Recurring (Non-one-time) Contributors}}\\
H2a & \textbf{PR volume$^\ast$ }     & 497.30  & +6.91\%  & 99.9\% & 0.69 & \spark{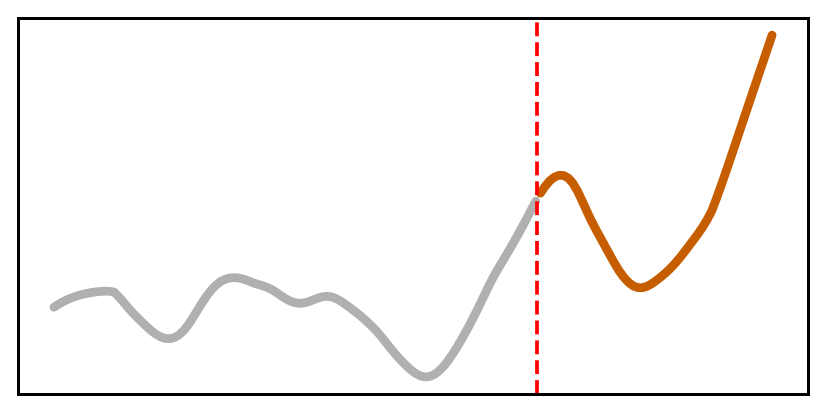}\\
H2b & \textbf{PR merge.$^\ast$ }    & -0.03  & -3.45\%  & 99.9\% & 2.14 & \spark{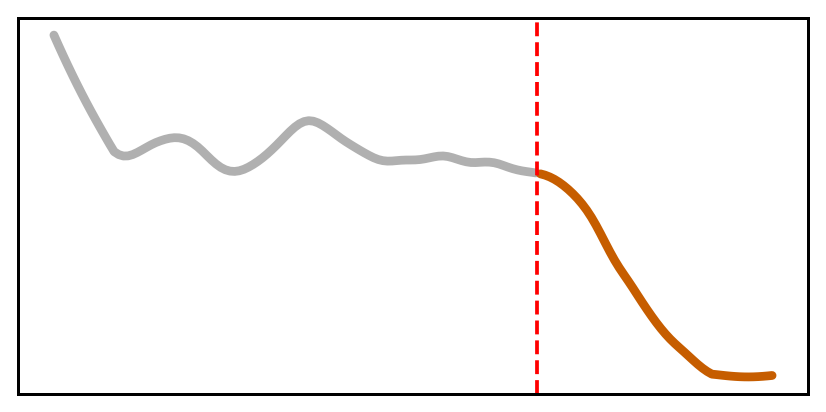}\\
H2c & Issue volume$^\ast$   & -453.11 & -12.40\% & 99.9\% & 0.97 & \spark{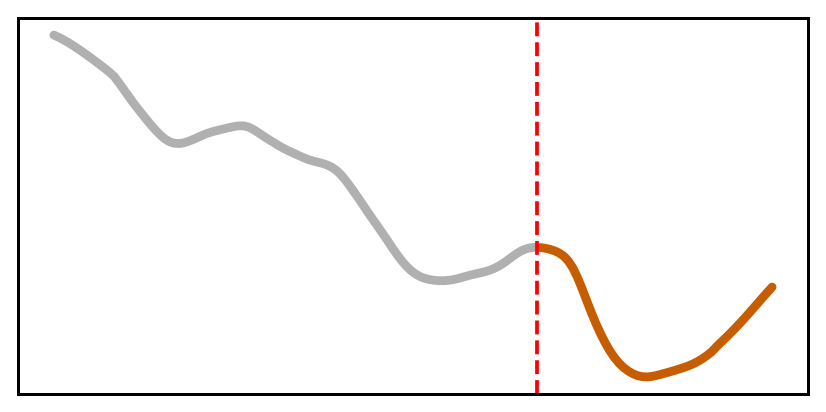}\\
H2d & \textbf{Issue compl.$^\ast$} & -0.05  & -8.09\%  & 99.9\% & 1.60 & \spark{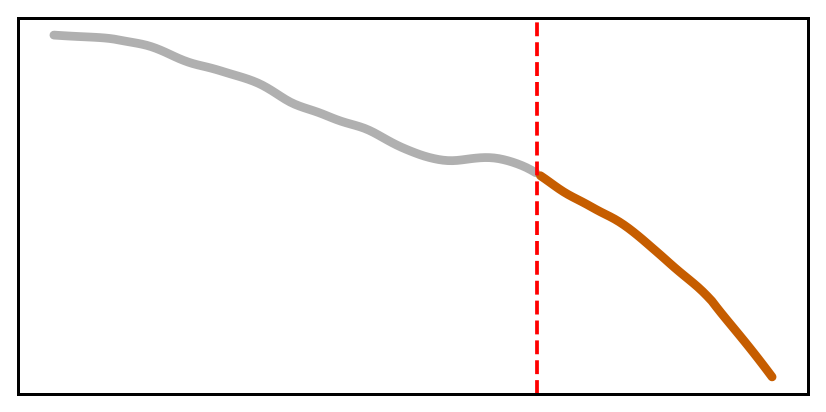}\\

\addlinespace[1pt]
\rowcolor{gray!12}\multicolumn{7}{@{}l}{\textit{Transient (One-time) Contributors}}\\
H3.a & \textbf{PR volume$^\ast$ }     & 24.67   & +5.84\%  & 99.4\% & 0.43& \spark{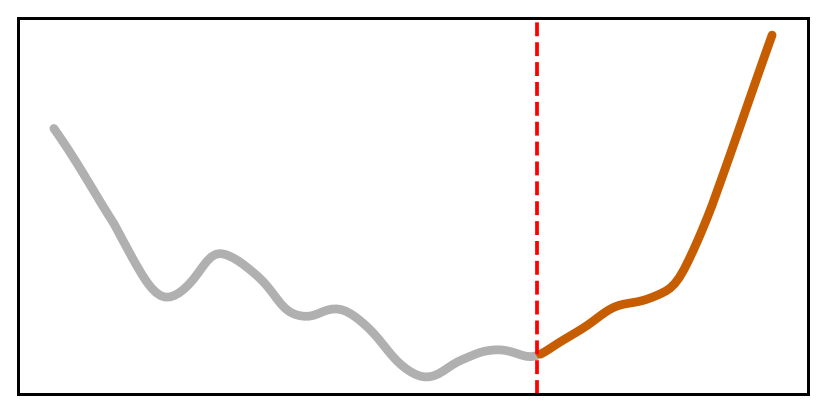}\\
H3.b & \textbf{PR merge.$^\ast$ }    & -0.09  & -18.18\% & 99.1\% & 2.84 & \spark{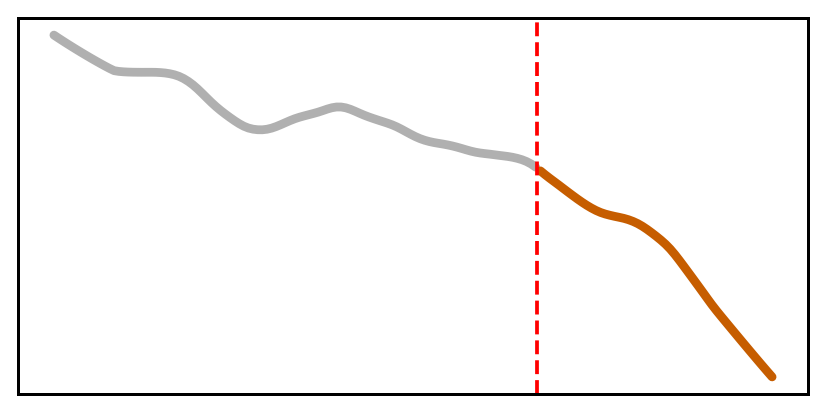}\\
H3.c & Issue volume          & 58.20   & +4.00\%  & 82.8\% & 0.27& \spark{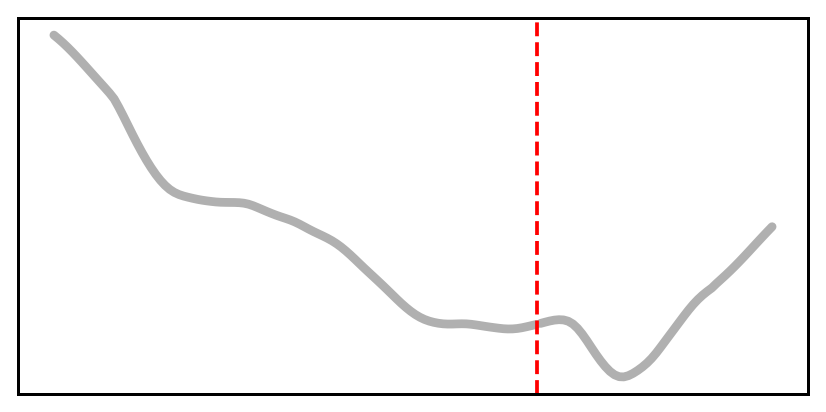}\\
H3.d & \textbf{Issue compl.$^\ast$} & -0.05  & -10.30\% & 99.9\% & 1.22 & \spark{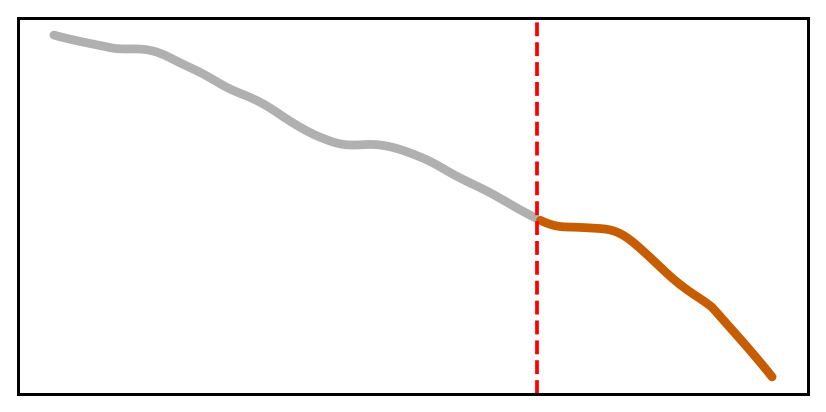}\\

\bottomrule
\end{tabular}
}

\vspace{2pt}
\begin{minipage}{\columnwidth}
\scriptsize
$^\ast$ Posterior probability exceeding 95\% indicates a statistically significant effect; hypotheses are bolded when observed effect are significant and in the hypothesized direction. Trend graphs show the pre-intervention period in \textcolor{gray!70!black}{gray}; significant post-intervention changes in \textcolor[HTML]{D55E00}{orange}; and non-significant changes in \textcolor{gray!70!black}{gray}. The dashed \textcolor{red!75!black}{red} line marks the AI-DDoS intervention. Hyp. = hypothesis, Abs. = absolute effect, Rel. = relative effect, Post. = posterior probability, and $|d|$ = Cohen's $d$.
\end{minipage}
\vspace{-18pt}
\end{table}

%% file: sections/4_RQ2.tex
\section{AI-DDoS Remediation Strategies (RQ2)}
\label{sec:rq2}

To answer \textbf{RQ2}, we continue following the PBR approach~\cite{lumineau2025roadmap} by engaging with OSS maintainers and practitioners as the primary stakeholders and using theoretical framing to investigate the remediation strategies that OSS projects use or plan to adopt in response to \aiddos.

\subsection{Method}

To identify these strategies, we conducted semi-structured interviews with 11 OSS maintainers, followed by a validation survey ($N=229$) to assess the perceived acceptability of interview-derived strategies across the OSS ecosystem.

\subsubsection{Interview Protocol}
We designed the interviews as a theory-informed follow-up to the AI-DDoS phenomena identified in RQ1. To examine how OSS communities respond to AI-DDoS as an organizational disruption, we used Organizational Resilience Theory (ORT)~\cite{vogus2007organizational, hillmann2021organizational} guiding our interview design and analysis. ORT is well suited here because it explains how organizations absorb disruption, adapt their practices, and learn from adversity \cite{musa2025absorptive}, providing an analytic lens to investigate the \textit{resilient behaviors} communities enact, the \textit{resources and capabilities} they mobilize, and the longer-term \textit{organizational growth} reflected in governance, review practices, contribution pathways, and resource allocation.

The interview contained five sections. We first collected background information about participants’ roles, project size, project domain, and maintainer team size, followed by an icebreaker question about their experiences with AI-generated contributions in OSS. The remaining sections were organized around ORT-informed constructs: \textit{Experience and Challenges}, which examined changes in contributor behavior, review work, and project coordination; \textit{Resilient Behaviors}, which captured how communities accepted, questioned, avoided, negotiated, or acted on AI-assisted contributions; \textit{Resilience Resources and Capabilities}, which examined how communities filtered, reviewed, coordinated, anticipated, and governed these contributions; and \textit{Outcomes and Growth}, which explored whether projects maintained review functions, recovered capacity, learned from disruption, and developed forward-looking strategies. The interview concluded with an open-ended question inviting participants to share any additional experiences or perspectives we may have missed. The details of the interview questions are provided in the supplementary document ~\cite{supply}.

\textbf{Sandbox and Pilot Study.} We refined the interview protocol through two internal sandbox sessions and two pilot interviews. The sandbox sessions helped improve question flow, wording, leading prompts, redundancy, and coverage. For example, we avoided the term ``AI-DDoS'' and used more neutral wording to ask about participants' experiences with AI-assisted contributions in OSS. Pilots with one OSS practitioner and one SE researcher led only to minor wording clarifications.

\textbf{Participant Recruitment.} We recruited 11 OSS maintainers through author contacts and snowball sampling~\cite{goodman1961snowball}, covering diverse project sizes, domains and governance structures. Eight identified as men and three as gender minorities; 81.82\% had more than four years of OSS experience and belonged to projects with over 100 contributors (details in~\cite{supply}).

The study was approved by our institution's IRB. Participants gave informed consent for participation and recording and were compensated with a \$50 Amazon gift card. Each 45--60 minute interview was conducted via video call with audio and screen recording. Transcripts were auto-generated and manually corrected by one author.

\subsubsection{Data Analysis}

To identify AI-DDoS remediation strategies, we conducted reflexive thematic analysis of the interview transcripts~\cite{braun2006using}, using an ORT analytic lens. Two authors first inductively open-coded governance actions, policies, tools, and practices that communities had adopted, considered, or rejected in response to AI-assisted contribution pressure. We then used ORT components, \textit{resilient behaviors}, \textit{resilience resources and capabilities}, and \textit{organizational growth}, to interpret how these actions helped communities absorb pressure, adapt workflows, or reconfigure longer-term contribution pathways. Through weekly negotiated agreements meetings, we refined codes, linked them to transcript segments, and iteratively adjusted the codebook. We then merged, split, and grouped related codes into higher-level strategy categories. We reached saturation after eight interviews, when no new codes emerged~\cite{francis2010adequate}, and conducted three additional interviews to confirm saturation and balance participant perspectives.

In total, we identified 11 remediation strategies and assessed their implementation effort as low, medium, or high based on participants' descriptions of required labor, tooling, workflow changes, and coordination. Guided by ORT, we then grouped the strategies into three resilience orientations: \emph{preservative} strategies protect existing review standards by absorbing pressure, \emph{adaptive} strategies modify current workflows to develop situation-specific responses, and \emph{transformative} strategies  reconfigure assumptions about contribution, trust, and participation \cite{lengnick2011developing} (Table \ref{tab:strategies}). Details are provided in the supplementary document~\cite{supply}.

\subsubsection{Member Checking}
We conducted member checking with interviewees to validate whether our interpretations accurately reflected their experiences, following established guidance~\cite{corbin2014basics}.  We sent each participant a summary of the synthesized strategies and invited feedback. Seven of the 11 participants responded, collectively covering all synthesized strategies. Their feedback confirmed our findings and provided minor clarifications, but yielded no new insights.

\subsubsection{Large Scale Validation Survey}
To validate the findings and assess community-level acceptance of identified strategies, we conducted a large-scale survey with 229 OSS practitioners. 

\textbf{Survey Design.} The survey comprised two sections. The first collected demographic and OSS background information (gender, years of OSS involvement, primary OSS role, and project size). The second presented the 11 strategies derived from interviews, asking respondents to rate their likelihood of adopting each in their projects on a 5-point Likert scale (1 = very unlikely, 5 = very likely). See details in \cite{supply}.

\textbf{Participant Recruitment.}  We recruited participants through our professional OSS networks and via email invitations to maintainers and contributors of the 294 repositories analyzed in RQ1, consistent with prior SE research involving OSS practitioners~\cite{feng2022case, feng2025domains}. The survey remained open for two weeks.

\input{Tables/strategies}


\textbf{Survey Responses.} We received 336 responses, of which 229 were valid. Participants represented diverse OSS roles, project sizes (fewer than 25 to more than 500 contributors), and experience levels, with most reporting 5--9 years (31.88\%) or 10--20 years (29.69\%) of OSS involvement. Of the participants, 197 identified as men (86.03\%) and 32 as gender minorities (13.97\%). See details in \cite{supply}. All participants provided informed consent under IRB-approved procedures.

For the open-ended questions, we had 83 substantive responses [S1-S83] that we qualitatively analyzed and found that no new mitigation strategies emerged. Most responses either expressed concerns about AI-DDoS or reinforced strategies already identified, e.g., \highlightquote{General limits on acceptable PR scope, i.e., outright refusing any large PRs that were not planned by the community in advance} [S39].


\subsection{Results}
We synthesized 11 strategies that OSS communities use or plan to use to respond to AI-DDoS pressure and grouped them into three resilience orientations informed by ORT \cite{vogus2007organizational, lengnick2011developing}:  \textit{Preservative}, \textit{Adaptive}, and \textit{Transformative} (Table~\ref{tab:strategies}). 


\subsubsection{Preservative Strategies are Holding the Line}

Across the three categories, endorsement is highest and most consistent for preservative strategies: tighten CI-based quality gates (SP1, 73\%) and enforce no-review cutoffs for red-flagged submissions (SP2, 71\%). Both are comparatively low-effort because they impose almost no new infrastructure or social cost. They route AI-assisted submissions through gates the project \emph{already} operates, such as continuous integration, tests, and existing scope/quality bars.  \highlightquote{If we do get huge, large submissions that drastically rewrite things, or don’t have approval, we will just auto-reject.} [P5]  \highlightquote{If it looks like it is AI-generated, we will just close the PR} [P6].  


This reflects a preservative resilience orientation in ORT: \textbf{communities first absorb disruption by adapting existing infrastructure to a new threat surface before pursuing deeper workflow changes}~\cite{lengnick2011developing}.


\subsubsection{Adaptive strategies draw conditional, divided support}
    
Adaptive strategies have mixed responses. Those  requiring low-to-medium effort were endorsed by more than half of respondents: require AI-use disclosure (SA1, 63\%), pre-review gates that clarify contributor intent or scope (SA2, 62\%), and flag risky submissions (SA3, 58\%). However, The endorsements are split for strategies that require high effort: Deploy AI reviewers as first-pass filters (SA4, 48\%), and expand triage through role rotation and community review (SA5, 34\%).

Communities are willing to adapt when changes are lightweight, such as a CONTRIBUTING.md file update [P7], PR template edit, or a heuristic label  (SA1--SA3). In contrast, they hesitate when adaptation depends on new tooling (SA4) or additional maintainer capacity (SA5). For example, AI reviewers (SA4) are adopted in some projects but not trusted to be autonomous: \highlightquote{LLMs are very poor at detecting other LLMs}~[P2]. Expand triage (SA5), the least endorsed adaptive strategy, can help distribute review work, but its feasibility depends on whether there are contributors with the time, expertise, and willingness to take on triage responsibilities.

These patterns reflect ORT’s resources-and-capabilities view: \textbf{\textit{communities adapt when changes are lightweight, but hesitate when they require new labor, unfamiliar tooling, or additional volunteer capacity~\cite{lengnick2011developing}}}.


\subsubsection{Transformative strategies change the object of governance, and divide along access}

Transformative strategies aim to transform the \emph{object of governance}. Endorsement decreases as these strategies become more effortful and restrictive.

The two most endorsed transformations carry medium effort and redefine \emph{what a contribution must demonstrate}. Verify contributor understanding before acceptance (ST1, 67\%) treats explanation and engagement as part of acceptance.  Aligning AI policy with project mission (ST2, 60\%) similarly ties governance to context: education-oriented projects may tolerate AI use \highlightquote{Because the project is education-oriented, I think in that sense, we end up being extremely lenient.}[P3], while security- or infrastructure-sensitive projects treat it more cautiously.

By contrast, the two least endorsed transformations require the highest effort because they introduce additional coordination and verification work for communities, especially maintainers. Prioritizing contributor intent in review (ST3, 43\%) moves evaluation upstream: contributors must explain what they are trying to change and why before code is reviewed, while maintainers must assess whether that intent is appropriate. As P5 explained, \highlightquote{now people have to provide their intent, what they're trying to fix, what's wrong, what they're changing, and we can just straight up say, hey, no.}~[P5]. Grounding trust in verified identity or endorsement (ST4, 39\%) moves governance even further, from the contribution to the contributor. As AI makes credible-looking submissions cheap to generate, some communities consider vouching for verification. \highlightquote{how do you make sure that someone who is contributing there is a human, is verifiable as a human?}~[P9].

The pattern reflects ORT's organizational growth, \textbf{\textit{communities support transformation when it improves contribution quality, but resist it when it needs high human effort}}~\cite{lengnick2011developing}.

%% file: Tables/strategies.tex
\begin{table*}[!htbp]
\caption{Identified 11 AI-DDoS remediation strategies with corresponding orientation, effort and validation evidences.}
\vspace{-4pt}
\centering
\resizebox{0.94\textwidth}{!}{
\begin{tabular}{p{3.8cm}p{9.5cm}ccc}

\toprule \toprule
\textbf{Strategy Name} & \textbf{Description} & 
\begin{tabular}[c]{@{}c@{}}\textbf{Member}\\\textbf{Checking}\end{tabular}
& \textbf{Effort} & \textbf{Responses\%}\\
\midrule

\rowcolor{gray!12}\multicolumn{5}{@{}l}{\textit{Preservative--- \textbf{Absorb} the pressure}} \\

\textbf{SP1.} Tighten CI-based quality gates
& \begin{tabular}[c]{@{}l@{}}Using existing CI, testing, and vulnerability checks to catch defects in\\AI-assisted submissions before maintainers spend review time.\end{tabular}
& \checkmark \checkmark \checkmark
& \adoptLow
& \likertCell{9}{6}{12}{30}{43}{73} \\

\addlinespace
\textbf{SP2.} Enforce no-review cutoffs for red-flagged submissions
& \begin{tabular}[c]{@{}l@{}}Closing submissions that trigger obvious red flags such as drastic rewrites, \\excessive file changes, missing approval, scope violations.\end{tabular}
&  \checkmark \checkmark
& \adoptLow
& \likertCell{6}{8}{15}{34}{37}{71} \\

\midrule \addlinespace
\rowcolor{gray!12}\multicolumn{5}{@{}l}{\textit{Adaptive--- \textbf{Accommodate} the pressure}} \\

\textbf{SA1.} Require AI-use disclosure
& \begin{tabular}[c]{@{}l@{}}Requiring contributors to state when AI tools were used, so maintainers can\\track provenance and treat undisclosed AI use as a process violation.\end{tabular}
& \checkmark
& \adoptLow
& \likertCell{8}{15}{14}{19}{44}{63} \\

\addlinespace
\textbf{SA2.} Add pre-review contribution gates
& \begin{tabular}[c]{@{}l@{}}Requiring contributors to clarify their intent, planned changes, or\\permissions before submissions reach maintainer review.\end{tabular}
& \checkmark \checkmark
& \adoptMid
& \likertCell{4}{11}{23}{38}{24}{62} \\

\addlinespace
\textbf{SA3.} Flag risky submissions through heuristics
& \begin{tabular}[c]{@{}l@{}} Using behavioral signals (e.g., account age, submission size, commit cadence,\\bursty PR patterns) to mark submissions for additional scrutiny during review.\end{tabular}
& \checkmark \checkmark
& \adoptMid
& \likertCell{9}{12}{21}{33}{25}{58} \\

\addlinespace
\textbf{SA4.} Deploy AI reviewers as first-pass filters
& \begin{tabular}[c]{@{}l@{}}Using AI-powered review tools (e.g., CodeRabbit, Greptile, Graphite) as a\\first-pass filter on incoming submissions, freeing reviewers for higher-value work.\end{tabular}
&  \checkmark \checkmark
& \adoptHigh
& \likertCell{23}{16}{13}{24}{24}{48} \\

\addlinespace
\textbf{SA5.} Expand triage through rotation \& community review
& \begin{tabular}[c]{@{}l@{}}Increasing triage capacity through rotating triage roles or inviting community\\members to help organize and review incoming submissions.\end{tabular}
& \checkmark \checkmark \checkmark \checkmark
& \adoptHigh
& \likertCell{15}{19}{32}{25}{9}{34} \\

\midrule
\rowcolor{gray!12}\multicolumn{5}{@{}l}{\textit{Transformative--- \textbf{Reconfigure} under the pressure}} \\

\textbf{ST1.} Verify contributor understanding before acceptance
& \begin{tabular}[c]{@{}l@{}}Requiring contributors to show that they understand what they submitted,\\through reflection, explanation, or demonstrated technical grounding.\end{tabular}
& \checkmark \checkmark \checkmark
& \adoptMid
& \likertCell{3}{10}{20}{41}{26}{67} \\

\addlinespace
\textbf{ST2.} Align AI policy with project mission
& \begin{tabular}[c]{@{}l@{}}Designing AI policies around the project's purpose, risks, and values,\\rather than adopting a universal AI rule.\end{tabular}
&  \checkmark \checkmark
& \adoptMid
& \likertCell{9}{10}{21}{33}{27}{60} \\

\addlinespace
\textbf{ST3.} Prioritize contributor intent in review
& \begin{tabular}[c]{@{}l@{}}Shifting the unit of review from code to intent, evaluating the contributor's\\plan, design, and understanding before (or instead of) the code itself.\end{tabular}
&  \checkmark 
& \adoptHigh
& \likertCell{12}{14}{31}{29}{14}{43} \\

\addlinespace
\textbf{ST4.} Ground contribution trust in verified identity / endorsement
& \begin{tabular}[c]{@{}l@{}}Establishing trust at the identity layer by verifying personhood or\\requiring endorsement by trusted members.\end{tabular}
& \checkmark \checkmark
& \adoptHigh
& \likertCell{8}{21}{32}{26}{13}{39} \\

\bottomrule \bottomrule
\end{tabular}
}

\smallskip
{\scriptsize\centering
\textcolor{lkVL}{\rule{1.8ex}{1.4ex}}~Very Likely\quad
\textcolor{lkL}{\rule{1.8ex}{1.4ex}}~Likely\quad
\textcolor{lkN}{\rule{1.8ex}{1.4ex}}~Neutral\quad
\textcolor{lkU}{\rule{1.8ex}{1.4ex}}~Unlikely\quad
\textcolor{lkVU}{\rule{1.8ex}{1.4ex}}~Very Unlikely\par}

\label{tab:strategies}

\vspace{-14pt}
\end{table*}

%% file: sections/6_Limitations.tex
\section{Threats to validity}
\label{sec:threats}

Our study relies on multiple sources of empirical evidence, combining qualitative and quantitative methods to characterize AI-DDoS. We discuss the main threats to validity below.

\textbf{Construct Validity.}
The RQ1 Stage~1 qualitative analysis (Section~\ref{sec:rq1_stage1}) may overrepresent visible, vocal, or highly affected practitioners. Therefore, we treat it as phenomenon identification rather than prevalence estimation, triangulating each theme across three independent sources and validating with repository-level behavior across 294 GitHub repositories.

For RQ2, participants' retrospective accounts may be affected by self-selection bias~\cite{marcus2005people, rogelberg2003profiling}. We mitigated this by using neutral interview questions and asking participants to ground responses in concrete examples. We also used ORT to structure data collection and analysis, and the validation survey revealed no additional strategy categories, suggesting reasonably comprehensive theoretical framing. To further reduce interpretive subjectivity~\cite{onwuegbuzie2007validity}, we validated our strategies through a survey of 229 OSS community members across diverse project sizes, roles, and experience levels--- a sample comparable to many empirical SE studies~\cite{russo2024navigating}.

\textbf{Internal Validity.}
Our BSTS analysis estimates the counterfactual trajectory each metric would have followed absent the intervention. Deviations from this trajectory are consistent with AI-DDoS effects but do not rule out other 2025 ecosystem-level changes, such as platform growth or broader shifts in OSS participation. Because we do not observe individual AI use directly, we use 2025 as a pre-specified detection window, supported by Stage~1 practitioner convergence, broader evidence that AI-assisted development became routine that year~\cite{StackOverflow2025survey, github2025octoverse}, and prior causal impact work~\cite{thorakkattle2022forecasting}. Therefore, we interpret results from 294 widely visible, active repositories as ecosystem-level evidence rather than direct measurements of individual AI use. As a model robustness check, placebo tests using January 2024 as a pseudo-intervention date strengthen confidence in PR findings, but significant placebo effects for issue metrics suggest that issue workflows were already changing before 2025, and we therefore interpret issue-side findings more cautiously~\cite{lenhart2021effects}.

\textbf{External Validity.}
To cover sustained public OSS activity and ensure that repository-level changes are measurable over time, our dataset covers popular, actively maintained projects with visible contribution pathways. However, the findings may not fully generalize to lower-traffic, less active, privately managed, or less newcomer-oriented repositories. Similarly, the survey captures the views of OSS participants who chose to respond, and while the sample spans diverse project sizes, roles, and experience levels, it may not represent all OSS communities equally. Future work should examine whether similar AI-DDoS dynamics appear in smaller, private, or less visible projects and across longer observational windows as AI-assisted development continues to diffuse.

%% file: sections/5_Discussion.tex
\section{Concluding Remarks}
Our results converge on a concerning picture: \aiddos is emerging as an ecosystem crisis, and communities may respond in ways that undermine their own sustainability and the ``openness'' that defines them. This is the temptation conventional DDoS defense warns against: we do not protect a service by turning it off~\cite{garber2000denial}. For OSS, the conceptual task is to interpret what that warning means for open participation under AI-generated contribution pressure. This follows the goal of PBR: to document an emerging phenomenon and conceptualize it into practice-relevant understanding~\cite{lumineau2025roadmap}. That understanding turns on how communities absorb disruption while staying functional, so we examine the question through ORT, as introduced in RQ2.

\begin{figure}[t]
\centering
\includegraphics[width=0.8\linewidth]{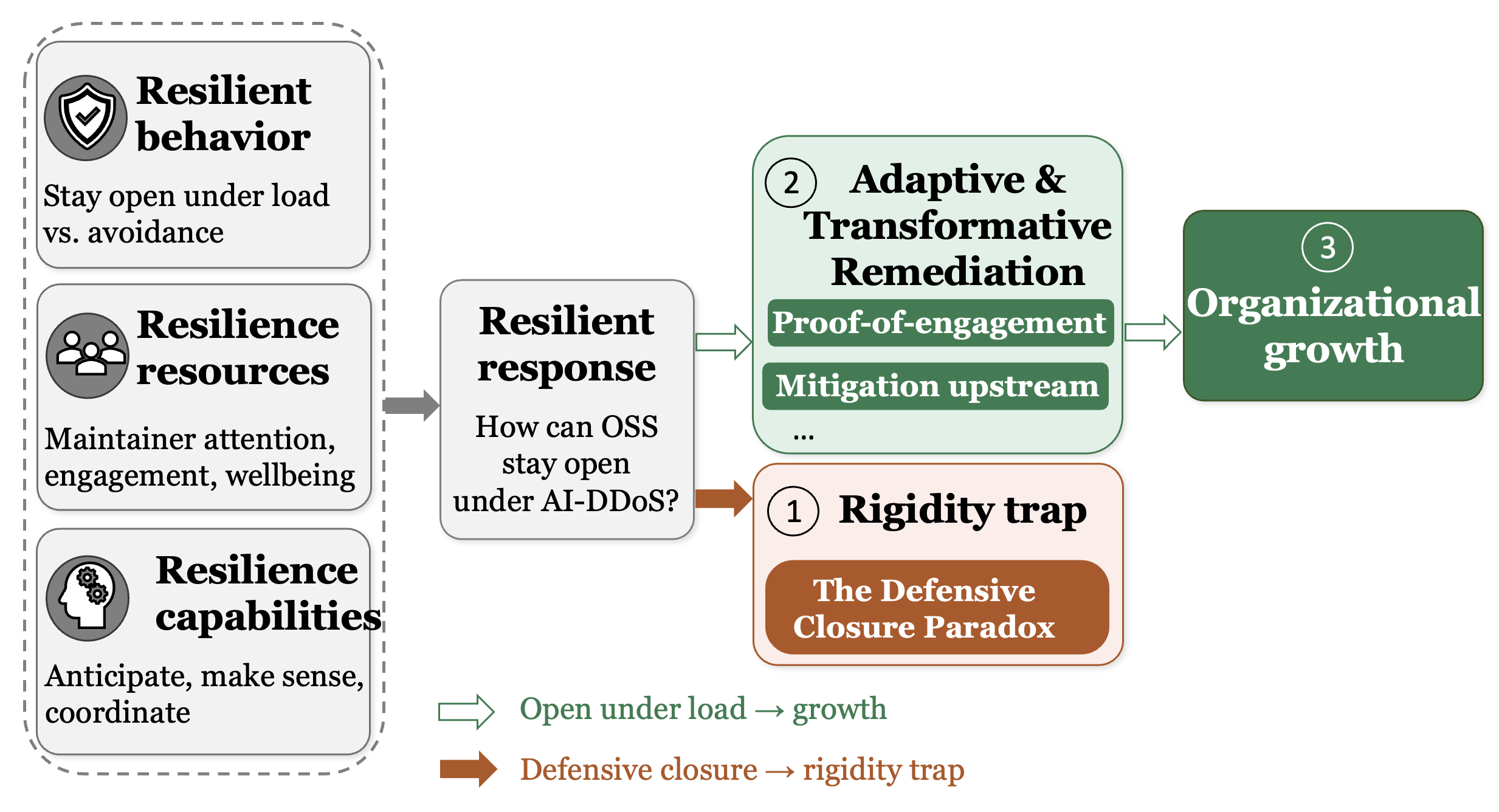}
\caption{AI-DDoS through organizational resilience theory.}
\label{fig:discussion}
\vspace{-14pt}
\end{figure}

As shown in Figure~\ref{fig:discussion}, ORT frames the central question through three capacities, \emph{resilient behavior}, \emph{resilience resources}, and \emph{resilience capabilities}: \textit{how can OSS stay open under \aiddos?} From there, the figure traces two paths. Defensive closure leads to a rigidity trap. Adaptive \& transformative remediation leads to organizational growth.

\textbf{Rigidity Trap: Defensive Closure.}
The red path (\circled{1} in Figure~\ref{fig:discussion}) shows defensive closure. Our survey responses (Table~\ref{tab:strategies}) show that OSS communities are already moving this way: participants widely endorsed preservative strategies because they are cheap, requiring little new infrastructure and little added coordination. The paradox is that these strategies protect maintainers' capacity in the short term by narrowing the contribution pathways that sustain OSS over the long term.

Such strategies can \highlightquote{kill most of the spray-and-pray submissions} [R3], but they may also fall hardest on the rough, first-attempt contributions that conventional good first issues were designed to welcome. Under AI-DDoS, what was a chronic depletion of maintainers' capacity~\cite{linaaker2024sustaining} becomes acute congestion, and closure starts to look rational: fewer entry points mean fewer submissions to review. But the damage is delayed. Because contributors take time to become long-term contributors~\cite{guizani2022attracting}, a project may appear stable while its future contributor pipeline erodes. The resulting rigidity trap is therefore more than a sustainability challenge for individual communities; it is a latent infrastructure risk for the broader software ecosystem that depends on them.

\textbf{Do Not Unplug the Server.}
If the rigidity trap is caused by treating closure as resilience, then the way out is to return to denial-of-service mitigation strategies~\cite{mahjabin2017survey, zargar2013survey}: keep the service available while separating legitimate from harmful traffic~\cite{mahjabin2017survey, yoon2010using}. In technical systems, this involves filtering and scrubbing traffic before it reaches the origin server, using challenge--response mechanisms that impose a small cost on each request~\cite{garber2000denial}, relying on reputation and allow-lists built from sender history~\cite{garber2000denial, yoon2010using}, balancing load across servers~\cite{belyaev2014towards}, and drawing on upstream actors large enough to absorb or deflect floods~\cite{somani2017scale}.

The adaptive and transformative strategies in Table~\ref{tab:strategies} map onto these mitigation strategies, as OSS contributions are social interactions. For example, preservative gates resemble origin-side filters: cheap, familiar, already deployed, but risky. AI reviewers and behavioral flags act as scrubbing layers that identify suspicious submissions before they consume maintainer attention. Verifying contributor understanding and gating on stated intent operate as challenge--response mechanisms by requiring contributors to demonstrate context before receiving maintainer attention. Identity and endorsement provide reputation signals. Triage rotation and community review distribute the review load across more reviewers. Cross-project reputation and identity verification require action at the platform level, which are upstream of any single project.

For example, \emph{proof of engagement} (\circled{2} in Figure~\ref{fig:discussion}) maps onto challenge--response: contributors are asked to explain why a change is needed, how the code works, what alternatives were considered or failed, and how they respond to review. Projects can design questions that are cheap for genuine newcomers who have engaged with the project but costly for volume contributors. AI can generate a plausible pull request, but situated understanding is harder to fake across the discussion, revision, and review cycle. For contributors, the path through the door is therefore to make their engagement with the project visible: disclose AI use, discuss the issue before opening a PR, explain the reasoning behind the change, and respond substantively to review.

\emph{Moving mitigation upstream} is another example. The pressure cannot be resolved by individual projects alone, especially when AI-DDoS is a structural ecosystem problem. It is produced not only by individual contributors misusing AI, but also from platform incentives that reward visible activity~\cite{rudra2026vouch}, such as contribution graphs, profile visibility, and pull-request counts, also drive it. AI lowers the cost of producing these signals at scale~\cite{ebbers2026contributions}, creating a collective-action mismatch: the activity is generated at platform scale, but remediation falls on maintainers one pull request at a time. If platforms benefit from contribution activity, they should also help restore the trust, reputation, and accountability infrastructure that individual projects cannot build alone.

\textbf{Organizational Growth: Rebuilding Sustainability Infrastructure.}
Organizational growth (\circled{3} in Figure~\ref{fig:discussion}) depends on the OSS ecosystem to transform the \aiddos pressure into shared sustainability infrastructure. Bringing this transformation about raises three questions for future research. 

First, \emph{lightweight proof-of-engagement ``gates"} will allow OSS communities to ask contributors for small signs of real engagement before maintainers invest review effort. Crossing these gates should be easy for genuine newcomers and harder for low-context, high-volume contributors. Future work should investigate which questions or tasks best reveal situated understanding, genuine engagement and interest in the project, and how quickly agentic tools learn to bypass these gates.

Second, \emph{portable trust signals across projects} could reduce the need to evaluate every contributor from scratch. Cross-project reputation, contributor history, and personhood verification may help recognize trustworthy participants earlier~\cite{steinmacher2019overcoming}. Future research should investigate how to design such trust signals without turning OSS into a closed ecosystem.

Third, \emph{tools that help maintainers gauge engagement} could move automation beyond artifact checks such as whether tests pass or a pull request follows a template. Under AI-DDoS, maintainers also need support for engagement signals, like whether contributors explain their changes, respond to review, and understand project context. This raises a tension. If the tools that infer engagement quality are themselves AI-based, and AI is what generates the contributions, then the signals these tools read can be faked as cheaply as the contributions they were meant to screen. Future research should explore how to surface interaction signals in ways that resist this collapse and keep human judgment central.

\textbf{In conclusion}, AI-assisted contributions are placing OSS communities under structural pressure. This is a sustainability trap as much as a volume problem. The cheapest ways to protect community capacity in the short term will narrow the open pathways built over decades. Filtering AI-assisted work is the easy part. 
The harder challenge is to identify hard-to-fake signals and rebuild the trust that openness depends on. As one practitioner put it, the community {\textit{needs to build new trust, or agents will completely solve it for us}} [S22].



